\documentstyle[12pt]{article}
\def\PBHC{\hat{\bar{\cal P}}}
\def\PHC{\hat{\cal P}}

\begin{document}

\thispagestyle{empty}

\baselineskip=0.6cm

\noindent P.~N.~Lebedev Institute Preprint     \hfill
FIAN/TD/10--93\\ I.~E.~Tamm Theory Department       \hfill
\begin{flushright}{May 1993}\end{flushright}

\begin{center}

\vspace{0.5in}

{\Large\bf UNIFIED CONSTRAINED DYNAMICS}

\bigskip

\vspace{0.3in}
{\large  I.~A.~Batalin and I.~V.~Tyutin}\\
\medskip  {\it Department of Theoretical Physics} \\ {\it  P.~N.~Lebedev
Physical Institute} \\ {\it Leninsky prospect, 53, 117 924, Moscow,
Russia}$^{\dagger}$\\

\end{center}

\vspace{1.5cm}

\centerline{\bf ABSTRACT}
\begin{quotation}

The unified constrained dynamics is formulated without making
use of the Dirac splitting of constraint classes. The strengthened,
completely--closed, version of the unified constraint algebra
generating equations is given. The fundamental phase variable
supercommutators are included into the unified algebra as well.
The truncated generating operator is defined to be nilpotent
in terms of which the Unitarizing Hamiltonian is constructed.

\end{quotation}
\vfill
\noindent
$^{\dagger}$ E-mail address: lirc@glas.apc.org

\newpage

\setcounter{page}{2}

\section{Introduction}
In previous papers \cite{1,2} of the present authors a unified algebraic
description of Hamiltonian constraints has been proposed. The main motivation
of the description is to avoid explicit splitting \cite{3} of constraints into
the first-- and second--class ones. With this purpose one should include the
fundamental phase variable commutators into the unified constraint
algebra as well. Thus we conclude that dynamically--passive ghosts
should be assigned to the fundamental phase variables.

A version of generating equations of the unified constraint algebra
has also been given in papers \cite{1,2}. These equations generate all
the required commutators indeed, but they appear to be too weak
to restrict their natural arbitrariness by the canonical
transformations only.

In the present paper we give a strengthened version of generating
equations that provides for their arbitrariness to be of the
canonical nature certainly. Then we construct the Unitarizing
Hamiltonian that governs the unified constrained dynamics in the
extended phase space.

Notations and Conventions. As is usual, $\varepsilon(A)$ and
$\hbox{gh}(A)$ stand, respectively, for the Grassmann parity and ghost
number of a quantity $A$. The standard supercommutator is denoted by :

$$[\hat{A},\hat{B}]\equiv\hat{A}\hat{B}-\hat{B}\hat{A}
(-1)^{\varepsilon(\hat{A})\varepsilon(\hat{B})}.$$

\section{Strengthened Version of Generating Equations}

Let :

$$\hat{\Gamma}^A,\quad A=1,\dots,N,\quad\varepsilon(\hat{\Gamma}^A)\equiv
\varepsilon_A,\quad\hbox{gh}(\hat{\Gamma}^A)=0,\eqno{(2.1)}$$
be an initial set of fundamental phase variable operators.

Following the papers \cite{1,2}, let us assign a dynamically--passive
ghost parameter $\Gamma^*_A$ to each operator $\hat{\Gamma}^A$ of the set
(2.1):

$$\hat{\Gamma}{}^A\quad\longmapsto\quad\Gamma^*_A,\quad
\varepsilon(\Gamma^*_A)=\varepsilon_A+1,\quad\hbox{gh}(\Gamma^*_A)=1.
\eqno{(2.2)}$$

The dynamical passivity implies that $[\Gamma^*_A,\Gamma^*_B]\equiv0$ and,
besides,
no ghost parameters $\Gamma^*_A$ have their own conjugate momenta.

In turn, let

$$\hat{\Theta}^\alpha(\hat{\Gamma}),\quad\alpha=1,\ldots,M<N,\quad
\varepsilon(\hat{\Theta}^\alpha)\equiv\varepsilon_\alpha,\eqno{(2.3)}$$
be a total set of operator--valued irreducible constraints. Let us
assign a pair of canonically--conjugated ghost operators to each
of the constraints (2.3) :

$$\hat{\Theta}{}^\alpha(\hat{\Gamma})\quad\longmapsto\quad
(\hat{C}_\alpha,\PBHC{}^\alpha),\quad\alpha=1,\ldots,M,
\eqno{(2.4)}$$

$$\varepsilon(\hat{C}_\alpha)=\varepsilon(\PBHC{}^\alpha)=\varepsilon_\alpha
+1,\quad\hbox{gh}(\hat{C}_\alpha)=-\hbox{gh}(\PBHC{}^\alpha)=1,\eqno{(2.5)}$$

$$(i\hbar)^{-1}[\hat{C}_\alpha,\PBHC{}^\beta]={\delta_\alpha}^\beta.
\eqno{(2.6)}$$

Next, let us introduce the operator functions :

$$\hat{\Omega}(\hat{\Gamma},\Gamma^*,\hat{C},\PBHC),\quad
\hat{\Delta}(\hat{\Gamma},\Gamma^*,\hat{C},\PBHC), \eqno{(2.7)}$$

$$\hat{\Omega}_\alpha(\hat{\Gamma},\Gamma^*,\hat{C},\PBHC),
\quad \alpha=1,\ldots,M, \eqno{(2.8)}$$

$$\varepsilon(\hat{\Omega})=1,\quad\hbox{gh}(\hat{\Omega})=1,\quad
\varepsilon(\hat{\Delta})=0,\quad\hbox{gh}(\hat{\Delta})=2, \eqno{(2.9)}$$

$$\varepsilon(\hat{\Omega}_\alpha)=\varepsilon_\alpha+1,
\quad\hbox{gh}(\hat{\Omega}_\alpha)=1, \eqno{(2.10)}$$
to satisfy the following equations :

$$(i\hbar)^{-1}[\hat{\Omega},\hat{\Omega}]=\hat{\Delta},\quad
\hat{\Delta}\left|_{{}_{\Gamma^*=0}}=0,\right. \eqno{(2.11a,b)}$$

$$(i\hbar)^{-1}[\hat{\Omega},\hat{\Omega}_\alpha]=0,\quad
(i\hbar)^{-1}[\hat{\Omega}_\alpha,\hat{\Omega}_\beta]=0
\eqno{(2.12a,b)}$$

These equations require their compatibility conditions :

$$(i\hbar)^{-1}[\hat{\Omega},\hat{\Delta}]=0,\quad
(i\hbar)^{-1}[\hat{\Delta},\hat{\Omega}_\alpha]=0. \eqno{(2.13a,b)}$$
to be fulfilled.

The previously--given version \cite{1,2} of generating equations
did not include the operators (2.8) so that the equations (2.12)
were also absent. Thus, in fact, we have been dealt with the
equations (2.11) only. However, being the ghost parameters $\Gamma^*_A$
different from zero, these equations themselves appear to be
insufficient to determine a certain solution for $\hat{\Omega}$, even
up to a canonical transformation. To fix such a solution, one
needs a given operator $\hat{\Delta}$ satisfying the compatibility condition
(2.13a) selfconsistently.

The present, strengthened, version includes the new equations (2.12)
that restrict effectively the arbitrariness of the operators $\hat{\Omega}$,
$\hat{\Delta}$, $\hat{\Omega}_\alpha$ and thus make it possible just determine
a solution for these operators modulo $\Gamma^*$--dependent canonical
transformataion in the extended phase space (2.1) $\oplus$ (2.4).

Given the constraints (2.3), one should seek for a solution to
the generating equations in the form of a series expansion in
powers of the ghost parameters $\Gamma^*$ and operators $\hat{C}_\alpha$,
$\PBHC{}^\alpha$ :

$$\hat{\Omega}=\hat{C}_\alpha\hat{\Theta}{}^\alpha(\hat{\Gamma})+
\ldots, \eqno{(2.14)}$$

$$\hat{\Delta}=-2\hat{C}_\alpha\Gamma^*_A\hat{E}^{A\alpha}(\hat{\Gamma})
(-1)^{\varepsilon_\alpha}-\Gamma^*_B\Gamma^*_A\hat{D}^{AB}(\hat{\Gamma})
(-1)^{\varepsilon_B}+\ldots, \eqno{(2.15)}$$

$$\hat{\Omega}_\alpha=\hat{C}_\beta\hat{\Lambda}^\beta_\alpha(\hat{\Gamma})+
\ldots . \eqno{(2.16)}$$

Classical counterparts of the corresponding operators should
satisfy the following rank conditions on the constraint surface :

$$\hbox{rank}\|\partial_A\Theta^\alpha(\Gamma)\|\bigl|_{{}_{\Theta=0}}=
\hbox{rank}\|\Lambda^\beta_\alpha(\Gamma)\|\bigl|_{{}_{\Theta=0}}=M,
\eqno{(2.17)}$$

$$\hbox{rank}\|E^{A\alpha}(\Gamma)\|\bigl|_{{}_{\Theta=0}}=M^\prime,
\eqno{(2.18)}$$

$$\hbox{corank}\|D^{AB}(\Gamma)\|\bigl|_{{}_{\Theta=0}}=M^{\prime\prime},
\quad M^\prime+M^{\prime\prime}=M. \eqno{(2.19)}$$

The conditions (2.18), (2.19) encode the presence of $M^\prime$ first--class
and $M^{\prime\prime}$ second--class constraints among the $M$
linearly--independent functions $\Theta^\alpha(\Gamma)$. That is the meaning
of the functions $E^{A\alpha}(\Gamma)$ and $D^{AB}(\Gamma)$.

As for the functions $\Lambda^\beta_\alpha(\Gamma)$, their meaning is that
the linear combinations
$\tilde{\Theta}^\alpha\equiv(\Lambda^{-1})^\alpha_\beta\Theta^\beta$ are
Abelian constraints. Of course, the inverse matrix $\Lambda^{-1}$ may appear
to be nonlocal in field--theoretic case so that explicit use of the Abelian
constraints $\tilde{\Theta}^\alpha$ themselves is rather undesirable in
general.

Substituting the expansions (2.14) -- (2.16) into the generating
equations (2.11), (2.12) and then solving these equations together
with their compatibility conditions (2.13) in all orders in ghosts,
one obtains in a usual way all the structural relations of the
unified constraint algebra. In particular, the fundamental
supercommutators $(i\hbar)^{-1}[\hat{\Gamma}^A,\hat{\Gamma}^B]$ and
constraint involution relations are generated, respectively, to the
$(\Gamma^*)^2$ and $(\hat{C})^2$--orders of the equation (2.11), as it has
been shown in papers \cite{1,2}.

\section{ Unitarizing Hamiltonian}

Having the generating equations (2.11), (2.12) solved for the operators
$\hat{\Omega}$, $\hat{\Delta}$, $\hat{\Omega}_\alpha$, we are in a position
to construct the Unitarizing Hamiltonian. With this purpose let us first
introduce the operator function :

$$\hat{\Phi}(\hat{\Gamma},\Gamma^*,\hat{C},\PBHC),\quad
\varepsilon(\hat{\Phi})=0,\quad\hbox{gh}(\hat{\Phi})=0,
\eqno{(3.1)}$$
to satisfy the equations :

$$(i\hbar)^{-1}[\hat{\Phi},\hat{\Omega}_\alpha]=\hat{\Omega}_\alpha,
\eqno{(3.2)}$$

$$\hat{\Phi}\bigl|_{{}_{\Gamma^*=0}}={1\over2}(\hat{C}_\alpha\PBHC{}^\alpha
(-1)^{\varepsilon_\alpha}-\PBHC{}^\alpha\hat{C}_\alpha)\equiv\hat{\Phi}_0.
\eqno{(3.3)}$$
The operator $\hat{\Phi}_0$ is nothing other but $\hat{C}\PBHC$--contribution
to the total ghost number operator.

Let us suppose, that the equations (3.2) are solved for the operator
$\hat{\Phi}$ to be searched in the form of a series expansion in powers of the
ghost parameters $\Gamma^*_A$ and operators $\hat{C}_\alpha$,
$\PBHC{}^\alpha$. Then we define the truncated operator $\hat{\Omega}_T$ by
the formula :

$$\hat{\Omega}_T\equiv(i\hbar)^{-1}[\hat{\Phi},\hat{\Omega}],\quad
\varepsilon(\hat{\Omega}_T)=1,\quad\hbox{gh}(\hat{\Omega}_T)=1.
\eqno{(3.4)}$$

By definition, this operator possesses the properties :

$$(i\hbar)^{-1}[\hat{\Omega}_T,\hat{\Omega}_T]=0,\quad
(i\hbar)^{-1}[\hat{\Phi},\hat{\Omega}_T]=\hat{\Omega}_T, \eqno{(3.5a,b)}$$

The mentioned properties make it quite natural define the truncated
Hamiltonian operator :

$$\hat{H}_T(\hat{\Gamma},\Gamma^*,\hat{C},\PBHC),\quad
\varepsilon(\hat{H}_T)=0,\quad\hbox{gh}(\hat{H}_T)=0, \eqno{(3.6)}$$
to satisfy the equations :

$$(i\hbar)^{-1}[\hat{H}_T,\hat{\Omega}_T]=0,\quad
(i\hbar)^{-1}[\hat{\Phi},\hat{H}_T]=0, \eqno{(3.7)}$$
to be solved in the form of a series expansion in powers of the ghost
parameters $\Gamma^*_A$ and operators $\hat{C}_\alpha$, $\PBHC{}^\alpha$ :

$$\hat{H}_T=\hat{H}_0(\hat{\Gamma})+\ldots, \eqno{(3.8)}$$
where $\hat{H}_0$ is the initial Hamiltonian of the theory.

At the present stage we have to introduce the dynamically--active
Lagrangian multipliers and antighosts:

$$(\hat{\lambda}_\alpha,\hat{\pi}^\alpha)\quad\longmapsto\quad(\PHC_\alpha,
\hat{\bar{C}}{}^\alpha),\quad\alpha=1,\ldots,M, \eqno{(3.9)}$$

$$\varepsilon(\hat{\lambda}_\alpha)=\varepsilon(\hat{\pi}^\alpha)=
\varepsilon_\alpha,\quad\varepsilon(\PHC_\alpha)=
\varepsilon(\hat{\bar{C}}{}^\alpha)=\varepsilon_\alpha+1, \eqno{(3.10)}$$

$$\hbox{gh}(\hat{\lambda}_\alpha)=-\hbox{gh}(\hat{\pi}^\alpha)=0,\quad
\hbox{gh}(\hat{\PHC}_\alpha)=-\hbox{gh}(\hat{\bar{C}}{}^\alpha)=1,
\eqno{(3.11)}$$

$$(i\hbar)^{-1}[\hat\lambda_\alpha,\hat\pi^\beta]=\delta^\beta_\alpha,
\quad(i\hbar)^{-1}[\PHC_\alpha,\hat{\bar C}{}^\beta]=\delta^\beta_\alpha.
\eqno{(3.12)}$$

Then we construct the total charge :

$$\hat{Q}=\hat{\Omega}_T+\PHC_\alpha\hat{\pi}^\alpha, \eqno{(3.13)}$$

$$(i\hbar)^{-1}[\hat{Q},\hat{Q}]=0,\quad\varepsilon(\hat{Q})=1,\quad
\hbox{gh}(\hat{Q})=1, \eqno{(3.14)}$$
in terms of which physical operators $\hat{\cal O}$ and physical states
$|\hbox{Phys}\rangle$  are defined \cite{4} :

$$(i\hbar)^{-1}[\hat{\cal O},\hat Q]=0,\quad\hat Q|\hbox{Phys}\rangle=0.
\eqno{(3.15)}$$

Finally, we construct the Unitarizing Hamiltonian of the theory :

$$\hat H_{complete}=\hat H_T+(i\hbar)^{-1}[\hat \Psi,\hat Q], \eqno{(3.16)}$$

$$\varepsilon(\hat\Psi)=1,\quad\hbox{gh}(\hat\Psi)=-1. \eqno{(3.17)}$$

Being the gauge Fermion $\hat\Psi$ chosen in the simplest form :

$$\hat\Psi=\hat\chi_\alpha(\hat\Gamma)\hat{\bar C}{}^\alpha+
\hat\lambda_\alpha\PBHC{}^\alpha, \eqno{(3.18)}$$
classical counterparts of the gauge operators $\hat\chi_\alpha$ should
satisfy the unitary--limit rank conditions :

$$\hbox{rank}\|\{\Theta^\alpha,\chi_\beta\}\|\bigl|_{{}_{\chi=0,\Theta=0}}=
M^\prime, \eqno{(3.19)}$$
where the Poisson bracket $\{\, ,\, \}$ is defined to be a classical
counterpart
of the supercommutator $(i\hbar)^{-1}[\, ,\, ]$, and $M^\prime$ enters the rank
condition (2.18).

In fact, after the nilpotency property (3.5a) is established, the
Unitarizing Hamiltonian (3.16) is constructed along the lines of Ref.\cite{5}.

As is usual, one can show physical matrix elements of physical operators
to be gauge--independent :

$$\delta_{\hat\Psi}\langle\hbox{Phys}2|\hat{\cal O}\left(\hat\Xi(t)\right)
|\hbox{Phys}1\rangle\equiv0, \eqno{(3.20)}$$
where $\hat\Xi(t)$ is a solution to the Heisenberg equations of motion with the
Hamiltonian (3.16).

To conclude this Section, the following remark is in order. We have
constructed the Unitarizing Hamiltonian at the general position with
respect to the ghost parameters $\Gamma^*_A$. Of course, as the ghost
parameters are dynamically--passive, one can choose the value $\Gamma^*_A=0$
to be quite sufficient for all pragmatic aims after the fundamental
supercommutators are established. However, it seems to be more
geometrically--natural to consider arbitrary values for $\Gamma^*_A$ as far
as the physical content of the theory does not depend on $\Gamma^*_A$. That is
an aspect of the general ghost--decoupling property of the BFV--construction
(see Ref.\cite{6} for a review).

Being the value $\Gamma^*_A=0$ still chosen, one has at this point :

$$\hat\Omega_T\bigl|_{{}_{\Gamma^*=0}}=
\hat\Omega\bigl|_{{}_{\Gamma^*=0}}. \eqno{(3.21)}$$
By making use of this relation one can confirm directly that in case
the rank of fundamental supercommutators is constant, the Hamiltonian
(3.16) certainly reproduces, at the functional integral level, the
standard BFV description in which the constraints are split into the
first--class and second--class ones \cite{7}.

\section{ General Structure of Generating Operators}

Let us now return to the generating equations (2.11), (2.12) to study their
natural arbitrariness and transformation properties.

First of all, it follows from the equations (2.12b) that

$$\hat\Omega_\alpha=\hat{\cal U}^{-1}\hat C_\alpha\hat{\cal U},\quad
\varepsilon(\hat{\cal U})=0,\quad\hbox{gh}(\hat{\cal U})=0, \eqno{(4.1)}$$
where $\hat{\cal U}(\hat\Gamma,\Gamma^*,\hat C,\PBHC)$ is an arbitrary
canonical
transformation.

Then the equations (2.12a) yield :

$$\hat\Omega=\hat{\cal U}^{-1}\bigl(\hat C_\alpha
\hat{\tilde\Theta}{}^\alpha(\hat{\tilde\Gamma})+\Gamma^*_A
\hat{\tilde\Gamma}{}^A\bigr)\hat{\cal U}, \eqno{(4.2)}$$
where $\hat{\tilde\Gamma}{}^A(\hat\Gamma)$ is a ghost-independent
reparametrization of the fundamental phase variable operators.

Due to the equations (2.11) we have :

$$(i\hbar)^{-1}[\hat{\tilde\Theta}{}^\alpha,\hat{\tilde\Theta}{}^\beta]=0,
\eqno{(4.3)}$$

$$\hat\Delta=\hat{\cal U}^{-1}\bigl(-2\hat C_\alpha\Gamma^*_A(i\hbar)^{-1}
[\hat{\tilde\Gamma}{}^A,\hat{\tilde\Theta}{}^\alpha](-1)^{\varepsilon_\alpha}-
\Gamma^*_B\Gamma^*_A(i\hbar)^{-1}[\hat{\tilde\Gamma}{}^A,
\hat{\tilde\Gamma}{}^B](-1)^{\varepsilon_B}\bigr)\hat{\cal U}. \eqno{(4.4)}$$
Thus we conclude that the operators $\hat{\tilde\Theta}{}^\alpha$ are Abelian
constraints.

Next, it follows from the equations (3.2), (3.3), (4.1) that

$$\hat\Phi=\hat{\cal U}^{-1}\hat\Phi_0\hat{\cal U}, \eqno{(4.5)}$$
and hence :

$$\hat\Omega_T=\hat{\cal U}^{-1}\hat C_\alpha\hat{\tilde\Theta}
{}^\alpha\hat{\cal U}.
\eqno{(4.6)}$$

Finally, the equations (3.7), (4.6) yield :

$$\hat H_T=\hat{\cal U}^{-1}\hat{\tilde H}_T\hat{\cal U}, \eqno{(4.7)}$$

$$(i\hbar)^{-1}[\hat C_\alpha\hat{\tilde\Theta}{}^\alpha,\hat{\tilde H}_T]=0,
\quad(i\hbar)^{-1}[\hat\Phi_0,\hat{\tilde  H}_T]=0. \eqno{(4.8a,b)}$$
It follows from the equation (4.8b) that the operator $\hat{\tilde H}_T$ does
not depend on $\Gamma^*_A$. As for the equation (4.8a), its natural
arbitrariness is :

$$\hat{\tilde H}_T\quad\rightarrow\quad\hat{\tilde H}_T+
(i\hbar)^{-1}[\hat K,\hat C_\alpha\hat{\tilde\Theta}{}^\alpha], \eqno{(4.9)}$$

$$\varepsilon(\hat K)=1,\quad(i\hbar)^{-1}[\hat\Phi_0,\hat K]=-\hat
K. \eqno{(4.10)}$$
By choosing in the expression (3.16) :

$$\hat\Psi=\hat{\cal U}^{-1}\hat{\tilde\Psi}\hat{\cal U}, \eqno{(4.11)}$$
where the operator $\hat{\tilde\Psi}$ is $\Gamma^*$--independent, we arrive at
the following representation :

$$\hat H_{complete}=\hat{\cal U}^{-1}\hat{\tilde H}_{complete}\hat{\cal U},
\eqno{(4.12)}$$

$$\hat H_{complete}\equiv\hat{\tilde H}+(i\hbar)^{-1}[\hat{\tilde\Psi},\bigl(
\hat C_\alpha\hat{\tilde\Theta}{}^\alpha+\PHC_\alpha\hat\pi^\alpha\bigr)].
\eqno{(4.13)}$$

Thus we conclude that the $\Gamma^*$--dependence of the Unitarizing
Hamiltonian is absorbed totally into the overall operator $\hat{\cal U}$ of an
arbitrary canonical transformation.

\section{ Conclusion}

So, we have formulated the unified constrained dynamics without making use of
the Dirac splitting of constraint classes. The selfconsistency of the
fundamental supercommutators is guaranteed by including them into the unified
constrained algebra. The corresponding algebra--generating equations
are shown to be able to determine their solution effectively up to
a canonical transformation in the extended phase space. The truncated
generating operator is then defined to be nilpotent, in terms of
which the Unitarizing Hamiltonian is constructed.

\vspace{0.5cm}

{\bf Acknowledgement.} This work was supported, in part,
by a Soros Foundation Grant awarded by the American Physical Society.


\newpage

\begin{thebibliography}{9}

\bibitem{1} I.~A.~Batalin and I.~V.~Tyutin, Nucl. Phys. {\bf B  381}
(1992) 619.

\bibitem{2} I.~A.~Batalin and I.~V.~Tyutin, Phys. Lett. {\bf B} (1993)

\bibitem{3} P.~A.~M.~Dirac, {\it Lectures on Quantum Mechanics}, Yeshiva
University (Academic Press, New York, 1964).

\bibitem{4} T.~Kugo and I.~Ojima, Phys. Lett. {\bf B 73} (1978) 459.

\bibitem{5} I.~A.~Batalin and E.S.Fradkin, Phys. Lett. {\bf B 128} (1983) 303.

\bibitem{6} M.~Henneaux, Phys. Rep. {\bf 126} (1985) 1.

\bibitem{7} E.~S.~Fradkin and T.~E.~Fradkina, Phys. Lett. {\bf B 72} (1978)
343.

\end{thebibliography}
\end{document}